\documentstyle[12pt]{article}
\def\beq{\begin{equation}}
\def\eeq{\end{equation}}
\def\bea{\begin{eqnarray}}
\def\eea{\end{eqnarray}}
\begin {document}
\begin{titlepage}
December 1995 \\
\begin{flushright}
HU Berlin-EP-95/27\\
\end{flushright}
\mbox{ }  \hfill hepth@xxx/9512023
\vspace{6ex}
\Large
\begin {center}
\bf{On scheme dependence of gravitational dressing of renormalization group
functions}
\end {center}
\large
\vspace{3ex}
\begin{center}
H. Dorn
\footnote{e-mail: dorn@ifh.de}
\end{center}
\normalsize
\it
\vspace{3ex}
\begin{center}
Humboldt--Universit\"at zu Berlin \\
Institut f\"ur Physik, Theorie der Elementarteilchen \\
Invalidenstra\ss e 110, D-10115 Berlin
\end{center}
\vspace{6 ex }
\rm
\begin{center}
\bf{Abstract}
\end{center}
It is shown that for 2D field theories only the first order coefficient of
the gravitationally dressed RG $\beta$-function is scheme independent.
This is valid even for matter theories with one dimensionless coupling,
where the first two coefficients of the original $\beta$-function are scheme
independent.
\vspace{3ex}
\end{titlepage}
\newpage
\setcounter{page}{1}
\pagestyle{plain}
\noindent
For the gravitational dressing of renormalization group $\beta$-functions of
two-dimensional field theories there has been established in $lowest$ order of
perturbation theory the remarkable universal formula \cite{Sch1,KKP,AG,TKS,GZ}
\beq
\bar{\beta}(g)~=~b(c)\cdot \beta (g)~.
\label{1}
\eeq
The factor $b(c)$ depends only on the central charge $c$ of the conformally
invariant theory around which one is perturbing
\beq
b(c)~=~\frac{2}{\alpha Q}~,~~~~~Q^2~=~\frac{25-c}{3}~,~~~~~\alpha ~=~
\frac{Q-\sqrt{Q^2-8}}{2}~.
\label{2}
\eeq
A simple argument \cite{D1} based on the $c$-theorem \cite{Z} shows that
(\ref{1}) cannot be valid to all orders. Therefore, it is an interesting
question to ask for the higher orders. But then in general one has to face the
problem of scheme dependence. A natural candidate to find a scheme
independent universal formula
\footnote{By universal we understand a formula which relates the coefficients
of the power series of $\bar{\beta}$ to that of $\beta $ and which requires
no information beyond $\beta $ and $c$.}
seemed to be theories with only one coupling constant. In the absence of a
linear term the first two coefficients of their $\beta $ are scheme
independent. A corresponding formula was derived in \cite{D1} having in mind
the
prejudice that scheme independence holds also for the corresponding dressed
$\bar{\beta}$ coefficients.

Meanwhile, there has been found \cite{Sch2} a second order formula based on a
completely different approach \cite{SchTs}, which differs from our result by a
factor 2 in a relative weight. One explanation of such a difference would be
scheme dependence of the dressing procedure \cite{D2}. Therefore, the first
aim of this note is to analyze from the very beginning the question of
scheme dependence within our own approach. The analysis of \cite{D1}
also involved manipulations with a divergent bare coupling. As a byproduct
of our improved treatment we will be able to avoid these formal steps.

Throughout this letter we use the notations and conventions of \cite{D1}.
In our discussion three $\beta $-functions appear. $\beta (g)$ denotes the
$\beta $-function of the original matter theory without gravity,
$\tilde{\beta}(g)$
is the $\beta $-function of the combined matter-gravity system refering
to scaling in the (unphysical) coordinate space. The condition
$\tilde{\beta}(g)=0$ fixes the gravitational dressing. Finally the
gravitationally dressed function $\bar{\beta}(g)$ is connected to the response
of the system to a
change in the cosmological constant which sets the physical scale after
coupling to quantum gravity.\\

Our starting point is the gravitationally dressed action
\bea
\tilde{S}&=&S_{c}~+~S_{L}~+~\sum_{i}g_{i}\int
\tilde{V}_{i}~\sqrt{\hat{g}}~d^{2}z~,
\nonumber \\
S_{L}[\phi \vert \hat{g}]&=&\frac{1}{8\pi}\int d^{2}z\sqrt{\hat{g}}
\Big (\hat{g} ^{mn}\partial _{m}\phi \partial _{n} \phi + Q\hat{R}\phi (z)
+m^{2} e^{\alpha \phi} \Big )~.
\label{3}
\eea
$S_c$ denotes the conformal matter action. The 2D metric is treated in
conformal gauge $g_{ab}=e^{\alpha \phi}~\hat{g}_{ab}$. The Liouville mass
parameter $m^2$ plays the role of the cosmological constant. For the relation
between
the dressed and undressed perturbation the ansatz
\beq
\tilde{V}(z)~=~e^{\delta \phi (z)}\cdot V(z)
\label{4}
\eeq
is made. The dimension of $\tilde{V}$ is $2-y$ with
\beq
y~=~\delta (\delta - Q)~.
\label{5}
\eeq
We extract the renormalization Z-factor needed to calculate $\tilde{\beta}$
from the two-point function of $\tilde{V}(z)$. $y>0$ acts as a regularization
parameter, effectively. From
\beq
\langle \tilde{V}(z_{1})\tilde{V}(z_{2})\rangle~=~
\langle \tilde{V}(z_{1})\tilde{V}(z_{2})\rangle _{0}~-~
g\int d^{2} z \langle \tilde{V}(z_{1})\tilde{V}(z_{2})\tilde{V}(z)\rangle
_{0}~+~O(g^{2})
\label{6}
\eeq
together with
\bea
\langle \tilde{V}(z_{1}) \tilde{V}(z_{2})\rangle _{0}&=&\left (\frac
{m^2}{\bar{\mu}^2}\right )^{\frac{Q-2\delta}{\alpha}}B_{2}(\delta)~
\vert z_{1}-z_{2} \vert ^{2y-4}~,
\nonumber \\
\langle \tilde{V}(z_{1})\tilde{V}(z_{2})\tilde{V}(z_{3})\rangle _{0}&=&
f\cdot \left ( \frac{m^2}{\bar{\mu}^2}\right )^{\frac{Q-3\delta}{\alpha}}B_{3}
(\delta ,\delta ,\delta )~(\vert z_{1}-z_{2}\vert
\vert z_{1}-z_{3}\vert \vert z_{2}-z_{3}\vert )^{y-2}
\label{7}
\eea
we get
\bea
\langle \tilde{V}(z_{1})\tilde{V}(z_{2})\rangle &=&
\left (\frac {m^2}{\bar{\mu}^2}\right )^{\frac{Q-2\delta}{\alpha}}
B_{2}(\delta)~\vert z_{1}-z_{2} \vert ^{2y-4}\nonumber \\
&-&
\frac{\pi g f B_{3}(\delta ,\delta ,\delta )}{\vert z_{1}-z_{2}\vert ^{4-3y}}
\left ( \frac{m^2}{\bar{\mu}^2}\right )^{\frac{Q-3\delta}{\alpha}}
\frac{\Gamma (1-y)(\Gamma (\frac{y}{2}))^{2}}{\Gamma (y)
(\Gamma (1-\frac{y}{2}))^{2}}~+~O(g^{2})~.
\label{8}
\eea
The constant $f$ parametrizes the 3-point function of the undressed $V$.
\footnote{$\bar{\mu}$ is a scale needed to form a dimensionless ratio.
For a complete analysis of all dimensionful parameters involved see
\cite{DO2}.}
\beq
A_N~=~\left (\frac{m^2}{\bar{\mu}^2}\right )^{\frac{Q-N\delta}{\alpha}}~B_N~,
{}~~~~~N=2,~3
\label{9}
\eeq
is the $z$-independent part of the 2- and 3-point function for exponentials
of the Liouville field $\phi $ \cite{DO,ZZ}. In contrast to \cite{D1} we keep
track
of the $m$ dependence introduced by the Liouville sector more carefully, $B_N$
contains no further $m$ dependence. After factorizing in (\ref{8}) an overall
factor $A_2$ we see that in the remaining part $m$ only appears in the
combination
$$ g~\left (\frac{m^2}{\bar{\mu}^2}\right )^{-\frac{\delta}{\alpha}}~. $$
Although the $N$-point functions of Liouville exponentials for $N>3$ have
not been constructed explicitly, the last statement on $m$ dependence
is valid for all $N$ \cite{DO2}.

To take care of the 2D coordinate space dimension of the bare coupling $g$
we introduce a RG-scale $\mu$ which has to be distinguished from $\bar{\mu}$
and define the dimensionless $Z$-factor for the coupling constant
renormalization by
\beq
g~\left (\frac{m^2}{\bar{\mu}^2}\right )^{-\frac{\delta}{\alpha}}~=~
\mu ^y~Z_g (y,g_r)~g_r~.
\label{10}
\eeq
This yields
\beq
\tilde{\beta}~=~\left.\mu \frac{\partial}{\partial\mu} g_{r}\right|
_{g,m,\bar{\mu},
y~\mbox{\footnotesize fix}}~=~\frac{-y~g_{r}}{1~+~g_{r}\frac{\partial \log
Z_{g}}
{\partial g_{r}}}~.
\label{12}
\eeq
Besides $Z_g$ the renormalization of (\ref{8}) involves the $Z$-factor of the
composite operator $\tilde{V}$
\beq
\tilde{V}~=~\mu ^{-y}\left ( \frac{m^2}{\bar{\mu} ^2} \right )^{-\frac{\delta}
{\alpha}}Z_{\tilde{V}}(y,g_r)\tilde{V}_r~.
\label{13}
\eeq
Due to $\int d^2 z \sqrt{\hat{g}}\tilde{V}_r=\frac{\partial \tilde{S}}
{\partial g_r}$ it is related to $Z_g$ by
\beq
Z_{\tilde{V}}^{-1}~=~Z_{g}~+~g_{r}\frac{\partial Z_{g}}{\partial g_{r}}~.
\label{14}
\eeq
Now the renormalized two-point function $\langle
\tilde{V}(z_1)\tilde{V}(z_2)\rangle _r~=~\mu ^{2y} \left(
\frac{m^2}{\bar{\mu}^2}\right )^{\frac{2\delta}{\alpha}}Z_{\tilde{V}}^{-2}
\langle \tilde {V}(z_1)\tilde{V}(z_2) \rangle$ expressed in terms of $g_r$
has to be finite. With (\ref{8}), (\ref{14}), (\ref{10}) this implies
\footnote{As discussed in \cite{D1} we use $B_3 /B_2 = 1+O(y)$.}
\beq
Z_g~=~1~+~\left (\frac{\pi f}{y}+\gamma\right )~g_r~+~O(g_r ^2)~.
\label{15}
\eeq
$\gamma $ is an arbitrary number. In contrast to \cite{D1} we do not restrict
ourselves to minimal subtraction with respect to the `y-regularization'.
{}From (\ref{12}) we get the corresponding $\beta$-function $\tilde{\beta}$
\beq
\tilde{\beta}(y,g_r)~=~-yg_r~+~(\pi f+\gamma y)g_r ^2~+~\tilde{\beta}_3 (y)
g_r ^3~+~...~.
\label{16}
\eeq
Of course the deviation from minimal subtraction parametrized by $\gamma$ has
no influence on the first two nonvanishing coefficients of the original matter
$\beta$-function $\beta(g_r )$ which is obtained for $y\rightarrow 0$
\beq
\beta _2~=~\tilde{\beta}_2 (0)~=~\pi f~,~~~~~\beta _3 ~=~\tilde{\beta}_3 (0)~.
\label{17}
\eeq
However, the freedom to choose $\gamma$ influences the dependence of $y$
on $g_r$ imposed by the condition $\tilde{\beta}(y,g_r)=0$. This condition
ensures independence of the unphysical scale in coordinate space and
fixes the gravitational dressing of the perturbation $V$ within our
ansatz (\ref{4})
\beq
\tilde{\beta}(y,g_r)=0~~~~~\Leftrightarrow ~~~~y=y(g_r)=\beta _2 g_r +
(\beta _3 +\beta _2 \gamma )g_r ^2+O(g_r ^3)~.
\label{18}
\eeq
Via (\ref{5}) this leads to
\beq
\delta ~=~\delta (g_r)~=~-~\frac{\beta _2}{Q}g_r~+~\frac{1}{Q}\left
(\frac{\beta _2 ^2}{Q^2}
-\beta _3 -\beta _2 \gamma \right )g_r ^2~+~O(g_r ^3)~.
\label{19}
\eeq

Before we proceed to the construction of the gravitational dressed
$\bar{\beta}$ we have to have a closer look on the condition $\tilde{\beta}=0$.
Although
for general $y>0$ the bare coupling is finite, eq. (\ref{12}) tells us that
$\tilde{\beta}=0$ at finite $y\neq 0$ can be achieved for the price
of infinite $Z_g$ i.e. infinite $g$ only. This effect is in complete analogy
to the situation at the nontrivial fixed point in $\Phi ^4$ theory in
$4-\epsilon$ dimensions \cite{BLZ}. There it has been understood and used to
derive
scaling relations not at but near the critical temperature for
distances $x$ very large (very small) compared to the lattice constant
(correlation length).

Let us look at (\ref{10}) in the form
\beq
gZ_g ^{-1}(y,g_r)\mu ^{-y}~=~\left (\frac{m^2}{\bar{\mu}^2}\right )
^\frac{\delta}{\alpha} g_r~.
\label{20}
\eeq
Although both $g$ and $Z_g$ diverge for $y\rightarrow y(g_r),~\delta
\rightarrow \delta (g_r)$ both sides of (\ref{20}) remain finite in this
limiting process. Since in addition the $m$-dependence in a factorized way is
completely
decoupled from the divergence just discussed, it is naturally to define
\beq
\bar{g}~=~ g Z_g ^{-1}(y,g_r)\mu ^{-y}
\label{21}
\eeq
as the bare coupling with respect to a change of $m$. Then
\beq
\bar{\beta}(g_r)~=~\left.m\frac{\partial}{\partial m}g_r \right| _{\bar{g},~
\bar{\mu}\mbox{
\footnotesize fix}}
\label{22}
\eeq
delivers us
\bea
\bar{\beta}(g_{r},\frac{m}{\bar{\mu}})&=&\frac{2\beta _2}{\alpha Q}~
g_{r}^{2}~+~\frac{2}{\alpha Q}\left(\beta _3+\gamma \beta _2 -\frac{\beta _2
 ^2}
{Q^{2}}\right)~g_{r}^{3}\nonumber \\ &+&\frac{4\beta _2 ^2}{\alpha
^{2}Q^{2}}~g_{r}^{3}~ \log \frac{m}{\bar{\mu}}~+~O(g_{r}^{4})~.
\label{23}
\eea
This coincides with \cite{D1} for $\gamma =0$.\\

In conclusion we want to stress that the origin of scheme dependence of second
order coefficients for the dressed RG-function $\bar{\beta}$ can be traced back
to the presence of a linear term in $\tilde{\beta}$. It does not spoil scheme
independence of the first two coefficients of the original $\beta$ since this
function is related to the $y\rightarrow 0$ limit where the linear term drops
out. However, the gravitational dressing fixed  by $\tilde{\beta}=0$ requires
$y\neq 0$. An alternative point of view would be to interpret $y$ or $\delta$
not as a regularization parameter but as a second coupling. In theories with
more than one coupling the second order contribution is scheme dependent.

As we have seen there is no universal formula relating
$\bar{\beta}$ to $\beta$ beyond lowest order. Nevertheless, if one lowers the
demands on universal validity, within a given renormalization scheme one can
look for expressions for the coefficients of $\bar{\beta}$ in terms of that of
$\beta$ which as additional information involve $c$ only. Then our formula
(\ref{23}) for $\gamma =0$ yields the answer for minimal subtracted
`y-regularization'. It would be interesting to derive a similar result in a
more standard
scheme as e.g. minimal subtracted dimensional regularization.

The divergence of the original bare coupling (related to the RG with respect
to the scaling in the unphysical coordinate space) in the presence of a
regularization, which is forced by $\tilde{\beta}(y,g_r)=0$, requires the
definition of a new bare coupling $\bar g$ as in (\ref{21}). The similarity
with
an effect in the theory of critical phenomena is remarkable.

A last comment concerns the $\log \frac{m}{\bar{\mu}}$ term in (\ref{23}).
In contrast to the $\beta$-function in the MOM-scheme in standard field
theories (see e.g. \cite{MOM}) our $m$-dependence exhibits no threshold
behaviour. We expect this to be related to the peculiar role of $m$ in
Liouville theory. A scaling of $m$ can be compensated by a shift of the
constant mode of the
Liouville field.
\\[5mm]
{\bf Acknowledgement}\\
I would like to thank J. Ambjorn, P. Di Vecchia, C. Schmidhuber, K. Sibold and
A.A. Tseytlin for discussions.

%%%%%%%%%%

\end{document}